\documentclass{article}
\usepackage{spconf,amsmath,graphicx}
\usepackage[hidelinks]{hyperref}
\usepackage{amssymb}
\usepackage{algorithm}
\usepackage{algpseudocode}

\usepackage{booktabs}
\usepackage{multirow}
\usepackage[table,xcdraw]{xcolor}
\usepackage{lipsum}
\usepackage{caption}

\newcommand\norm[1]{\left\lVert#1\right\rVert}

\def\L{{\cal L}}

\title{Two-Step Knowledge Distillation for Tiny Speech Enhancement}
\name{Rayan Daod Nathoo\sthanks{These authors contributed equally to this work.}, Mikolaj Kegler$^*$, Marko Stamenovic}
\address{Bose Corporation, USA}

\begin{document}
\ninept

\maketitle

\begin{abstract}

Tiny, causal models are crucial for embedded audio machine learning applications. Model compression can be achieved via distilling knowledge from a large teacher into a smaller student model. In this work, we propose a novel two-step approach for tiny speech enhancement model distillation. In contrast to the standard approach of a weighted mixture of distillation and supervised losses, we firstly pre-train the student using only the knowledge distillation (KD) objective, after which we switch to a fully supervised training regime. We also propose a novel fine-grained similarity-preserving KD loss, which aims to match the student's intra-activation Gram matrices to that of the teacher. Our method demonstrates broad improvements, but particularly shines in adverse conditions including high compression and low signal to noise ratios (SNR), yielding signal to distortion ratio gains of 0.9 dB and 1.1 dB, respectively, at -5 dB input SNR and 63$\times$ compression compared to baseline.

\end{abstract}

\section{Introduction}
\label{sec:introduction}

In recent years, deep neural network (DNN) models have become a common approach to the speech enhancement (SE) problem, due to their performance~\cite{msdns,wang2022stft,irvin2022selfsupervised}. However, large, powerful models are often unsuitable for resource-constrained platforms, like hearing aids or wearables, because of their memory footprint, computational latency, and power consumption~\cite{wang2022stft,chen2019deep,fedorov2020tinylstms,stamenovic2021weight}. To meet these constraints, audio TinyML research tends to focus on designing model architectures with small numbers of parameters, using model compression techniques to reduce the size of large models, or both~\cite{chen2019deep, fedorov2020tinylstms,stamenovic2021weight,cruse}.

Pruning is a popular method for reducing the size of DNN models for SE~\cite{chen2019deep,fedorov2020tinylstms,stamenovic2021weight,prune_quantize_distill}. The goal of pruning is to remove weights least contributing to model performance. In its simplest form, this can be performed post-training by removing weights with the lowest magnitudes. Online pruning, where the model is trained and pruned concurrently, builds on post-training pruning by exposing the model to pruning errors while training, allowing it to adapt to this form of compression noise~\cite{chen2019deep}. Unstructured pruning of individual weights can yield impressive model size reduction with little performance sacrifice, but corresponding savings in computational throughput are not possible without hardware support for sparse inference, which is unusual in embedded hardware. Structured pruning of blocks of weights and/or neurons is often designed with broader hardware compatibility in mind, but the performance drop tends to be larger than for the unstructured case~\cite{stamenovic2021weight}. 

In contrast to pruning, knowledge distillation (KD) adopts a different framework. 
The goal of KD is to utilize a strong pre-trained \textit{teacher} model to guide the training of the smaller \textit{student}~\cite{hinton, survey, categories}. The underlying assumption is that the pre-trained teacher network offers additional useful context compared to the ground truth data by itself. Unlike pruning, this process does not involve modifying the student network from its original \textit{dense} form, which reduces the complexity of the deployment process. 
In this work, we focus on KD due to its above-outlined benefits over pruning. 

KD methods have been applied to various classification tasks in the audio domain~\cite{schmid2022knowledge,yoon2022interkd}. However, KD has not been extensively explored for causal low-latency SE, which often requires tiny networks (sub-100k parameters) optimized for low-resource wearable devices, such as hearing aids~\cite{fedorov2020tinylstms,stamenovic2021weight}. So-called 
\textit{response-based} KD approaches use the pre-trained teacher model's outputs to train a student network~\cite{sub_band,response_distill_SE}. However, distillation can be further facilitated by taking advantage of intermediate representations of the two models, not just their outputs~\cite{survey}. One common obstacle in such 
\textit{feature-based} KD is the dimensionality mismatch between teacher and student activations due to the model size difference. To alleviate this issue,~\cite {fitnets} proposed aligning intermediate features, while~\cite{attention} used attention maps to do so. The latter was applied in the context of SE in~\cite{multi_view_attention} using considerably large, non-causal student models intended for offline applications. In~\cite{cross_similarity}, the authors addressed the dimensionality mismatch problem for the causal SE models by using frame-level Similarity Preserving KD~\cite{sp} (SPKD). SPKD captures the similarity patterns between network activations for different training examples and aims to match those patterns between the student and the frozen pre-trained teacher models. The authors of~\cite{cross_similarity} also introduced fusion blocks, analogous to~\cite{knowledge_review}, to distill relationships between consecutive layers.

Here, we show that the efficacy of conventional KD methods is limited for tiny, causal SE models. To improve distillation efficacy, we first extend the method from~\cite{cross_similarity} by computing SPKD for each bin of the latent representations, corresponding to the time frame (as in~\cite{cross_similarity}) but also the frequency bin of the input, thus providing more resolution for KD loss optimization. The proposed extension outperforms other similarity-based KD methods which we also explore. Second, we hypothesize that matching a large teacher model might be challenging for small student models and thus may lead to suboptimal performance. Inspired by~\cite{gift}, we propose a novel two-step framework for distilling tiny SE models. In the first step, the student is pre-trained using only the KD criterion to match the activation patterns of the teacher, with no additional ground truth supervision. The goal of this unsupervised KD pre-training is to make the student similar to the teacher prior to the main training. Then, the pre-trained student model is further optimized in a supervised fashion and/or using KD routines. We find that pre-training using the proposed SPKD method at the level of the individual bin of the latent representation, followed by fully supervised training yields superior performance compared to other distillation approaches utilizing weighted mixtures of KD and supervised losses. We report the performance of our method across various student model sizes, input mixture signal-to-noise ratios (SNRs), and finally, assess the similarity between the activation patterns of the teacher and distilled student.

\begin{figure*}[t]
    \centering\includegraphics[width=0.978\textwidth]{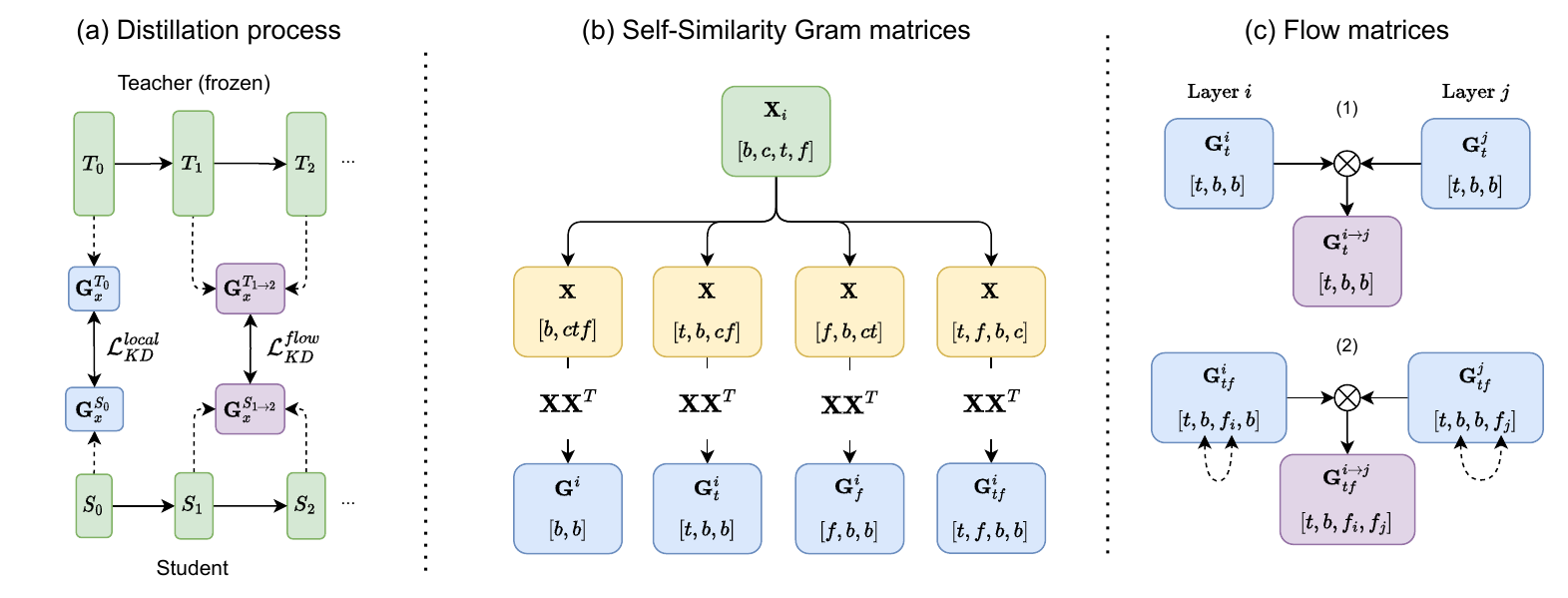}
    \caption{(a) Distillation process overview (b) Self-Similarity Gram matrices computation. (c) Flow matrices computation ($\bigotimes$ denotes matrix multiplication). Note that, for clarity, transpositions and matrix multiplications are applied only to the last two dimensions of each tensor.}
    \label{fig:bigfig}
\end{figure*}

\section{Methods}
\label{sec:methods}

\subsection{Model architecture}
\label{ssec:architecture}

Our backbone architecture for the exploration of tiny SE KD is the Convolutional Recurrent U-Net for SE (CRUSE) topology \cite{cruse}. However, note that the methodology developed here can, in principle, be applied to any other architecture. The CRUSE model operates in the time-frequency domain and takes power-law compressed log-mel spectrograms (LMS) as input. The LMS is obtained by taking the magnitude of the complex short-time Fourier transform (STFT, 512/256 samples frame/hop size), processing it through a Mel-space filterbank (80 bins, covering 50-8k~Hz range) and finally compressing the result by raising it to the power of $0.3$. The model output is a real-valued time-frequency mask bounded within the range $(0,1)$ through the sigmoid activation of the final block. The mask is applied to the noisy model input and reconstituted into the time domain using the inverse STFT and the noisy input phase.

The model comprises four encoder/decoder blocks and a bottleneck with grouped GRU units (4 groups), reducing the computational complexity compared to a conventional GRU layer with the same number of units~\cite{tan2019learning}. The encoder/decoder blocks are composed of 2D convolution/transpose convolution layers with (2, 3) kernels (time, frequency) and (1, 2) strides, followed by cumulative layer normalization~\cite{cLN} and leaky ReLU activation ($\alpha$ = 0.2). To further reduce the model complexity, skip connections between the encoder and decoder used in classic U-Net are replaced with 1x1 convolutions, whose outputs are summed into the decoder inputs. We enforce the model's frame-level causality by using causal convolutions and causal cumulative layer norms. The total algorithmic latency of the model is 32 ms (single STFT frame)~\cite{wang2022stft}.

In our experiments, both teacher and student are CRUSE models and their sizes are adjusted by changing the number of units in each block. In particular, the teacher uses \{32, 64, 128, 192\} encoder/decoder channels and 960 bottleneck GRU units, resulting in 1.9M parameters, and 13.34  MOps/frame (i.e. the number of operations required to process a single STFT frame). Our default student uses \{8, 16, 32, 32\} encoder/decoder channels and 160 bottleneck GRU units resulting in 62k parameters (3.3\% of the teacher), and 0.84 MOps/frame (6.3\% of the teacher).

\subsection{Self-similarity local knowledge distillation}
\label{ssec:local_kd}

Inspired by previous work~\cite{cross_similarity,sp}, we address the issue of dimensionality mismatch between teacher and student models by computing similarity-based distillation losses. The method captures and compares 
the relationship between batch items at each layer output, between teacher and student~(Fig.~\ref{fig:bigfig}a, $\L_{KD}^{local}$). We refer to this relationship as the self-similarity Gram matrix $\textbf{G}_x$.

Self-similarity matrices (Fig.~\ref{fig:bigfig}b) can be computed for an example network latent activation $\textbf{X}$ of shape [$b$, $c$, $t$, $f$], where $b$ - batch size, $c$ - channel, $t$ - activation \textit{width} (corresponding to the input time dimension), $f$ - activation \textit{height} (corresponding to the input frequency dimension), as shown in Fig.~\ref{fig:bigfig}b. The original implementation from~\cite{sp} involves reshaping $\textbf{X}$ to [$b$, $ctf$] and matrix multiplying it by its transpose $\textbf{X}^T$ to obtain the [$b$, $b$] symmetric self-similarity matrix $\textbf{G}$. Analogously, this operation can be performed for each $t$ or $f$ dimension independently with resulting $\textbf{G}_{t/f}$ matrices of size [$t/f$, $b$, $b$]. Such an increase in granularity improved the KD performance in~\cite{cross_similarity}. Here, we obtain even more detailed intra-activation Gram matrices by considering each $(t,f)$ bin separately, resulting in the $\textbf{G}_{tf}$ self-similarity matrix with shape~[$t$, $f$, $b$, $b$].

Finally, the local KD loss is computed using self-similarity matrices $\textbf{G}_x$ of any kind $x$ obtained from teacher $T$ and student $S$ as:
\begin{equation}
\label{eq:L_KD}
    \L_{KD}^{local} = \frac{1}{b^2} \sum_i \norm{\textbf{G}_x^{T_i} - \textbf{G}_x^{S_i}}^2_F,
\end{equation}
\noindent where $i$ is the layer index and $\norm{}^2_F$ is the Frobenius $l_2$ norm.

\subsection{Information flow knowledge distillation}
\label{sec:flow}

The above-outlined local similarity losses can be extended to capture relationships between activations of subsequent layers of the teacher and student models (Fig.~\ref{fig:bigfig}a, $\L_{KD}^{flow}$). The method is inspired by the Flow of Solution Procedure (FSP) matrices introduced in~\cite{gift} and aims to not only match local similarity between the teacher and student in the corresponding layers but also global inter-layer relations.
 
We propose two versions of flow matrices between layers $i$ and $j$ in our model (Fig.~\ref{fig:bigfig}c). The first one, $\textbf{G}_{t}^{i\rightarrow j}$, leverages $\textbf{G}_t$ self-similarity matrices. Thereby each self-similarity block shares the $t$-dimension and thus the interaction between the layers' self-similarity can be captured by performing matrix multiplication of $\textbf{G}_t^i$ and transposed $\textbf{G}_t^j$ (both sized [$t$, $b$, $b$]) for each time frame $t$.

The second version leverages $\textbf{G}_{tf}$ self-similarity matrices. However, the $f$ dimension in our model changes for each block due to the strided convolutions. To quantify the relationship between layers $i$ and $j$ of different dimensions we reshaped $\textbf{G}_{tf}^{i/j}$ to the size of [$t$, $b$, $f_{i/j}$, $b$]. Then for each time-batch-item pair ($t$,$b$), we obtain a [$f_{i/j}$, $b$] sub-matrix, which can be matrix multiplied with its transpose to obtain the flow matrix $\textbf{G}_{tf}^{i\rightarrow j}$ of size [$t$, $b$, $f_{i}$, $f_{j}$].

We define the loss similarly to Eq.~\ref{eq:L_KD} by  comparing the teacher $\textbf{G}^{T_{i\rightarrow j}}_x$ flow matrix with the student $\textbf{G}^{S_{i\rightarrow j}}_x$ flow matrix, of the same kind $x$, for every 2-layer-combination ($i$, $j$):

\begin{equation}
    \L_{KD}^{flow} = \frac{1}{b^2} \sum_{i} \sum_{j > i} \norm{\textbf{G}^{T_{i\rightarrow j}}_x - \textbf{G}^{S_{i\rightarrow j}}_x}^2_F
\end{equation}

\subsection{Training objective and two-step KD}
\label{sec:two_step}

We use phase-sensitive spectrum approximation (PSA)~\cite{erdogan2015phase}, with clean speech as a target, as the supervised portion $\L_{PSA}$ of the total loss. The KD portion $\L_{KD}$ of the total loss does not use the ground truth objective but instead, features obtained from the pre-trained, frozen teacher model. In particular, $\L_{KD}$ can match model outputs (i.e.~response distillation, analogous to~\cite{response_distill_SE}), $\textbf{G}_{x}$ ($\L_{KD}^{local}$), or $\textbf{G}^{i\rightarrow j}_{x}$ ($\L_{KD}^{flow}$) matrices introduced in Sections~\ref{ssec:local_kd} and~\ref{sec:flow}, respectively. $\L_{PSA}$ and $\L_{KD}$ are mixed using $\gamma$ coefficient to form the total loss:
\begin{equation}
\label{eq:L}
    \L = \gamma \L_{KD} + (1-\gamma) \L_{PSA}
\end{equation}
\noindent Inspired by~\cite{gift}, we propose a two-step KD approach by separating the student distillation process into two distinct parts. \textbf{Step 1}:~In the first step, $\gamma = 1$ for a fixed set of epochs to solely minimize $\L_{KD}$. While excluding the supervised $\L_{PSA}$ does not contribute to the optimal objective performance, this step provides strong initial weights for further student model training. \textbf{Step 2}:~After this pre-training step, the student is further optimized to maximize its objective performance using a fully supervised loss by setting $\gamma=0$, ($\L_{PSA}$ only) or a weighted $\L_{KD}$/$\L_{PSA}$ loss obtained by setting $\gamma =0.5$. For one-step KD using a weighted $\L_{KD}$/$\L_{PSA}$ loss, we set $\gamma=0.5$.

\subsection{Training setup}
During training, each epoch consists of 5,000 training steps, with each step being a 32-item batch of 2-second-long audio clips. We used the Adam optimizer with $6 \cdot 10^{-5}$ learning rate. The teacher model is trained till convergence to ensure the best performance for subsequent distillations. We train each student model for a total of 400 epochs (2M steps, 35.5k+ hours of audio). For student model pre-training (Step 1, Section~\ref{sec:two_step}), we use 100 epochs with $\gamma=1$, thus excluding supervised term $\L_{PSA}$ (Eq.~\ref{eq:L}).

\section{Results}

We use the dataset from the Interspeech 2020 Deep Noise Suppression (MS-DNS) Challenge~\cite{msdns} for experimentation, which consists of 500+ hours of clean speech and 100+ hours of noise, all mono clips sampled at 16 kHz. For model training, we mix the speech and noise at various SNR levels, sampled from a uniform distribution between -5 and 15 dB. We employ a LUFS-based SNR calculation for more perceptually relevant mixtures and to de-emphasize the effects of impulsive noises~\cite{lufs}. To evaluate the trained models we use the non-reverberant evaluation set consisting of 150 clips of noisy speech samples and their respective clean references.

We quantify the performance of each model via Signal-to-Distortion ratio~(SDR)~\cite{sdr}, wide-band Perceptual Evaluation of Speech Quality~(PESQ)~\cite{pesq}, and extended Short-Time Objective Intelligibility~(eSTOI)~\cite{jensen2016algorithm}. We also report scores obtained from DNS-MOS~P.835~\cite{dnsmos}, a DNN mean opinion score (MOS) estimator showing a good correlation with subjective ratings. All of our results are reported as improvements over unprocessed noisy inputs ($\Delta$).

\label{sec:results}
\begingroup
\setlength{\tabcolsep}{4pt} 
\begin{table}[]
\caption{One-step KD for tiny SE. \textit{Output}:~$\L_{KD}$ comparing teacher and student outputs (similar to~\cite{response_distill_SE}). $\textbf{G}_x$: Feature-based $\L_{KD}$ using self-similarity matrix of type $x$ (Fig.~\ref{fig:bigfig}b). All models are initialized with the same random weights and use $\gamma=0.5$ (Eq.~\ref{eq:L}).}\vspace{-4.5pt}
\label{tab:local_distill_table}
\resizebox{\columnwidth}{!}{
\centering
\small
\begin{tabular}{@{}ccccccc@{}}
\toprule
\rowcolor[HTML]{FFFFFF} 
\cellcolor[HTML]{FFFFFF} &
  \cellcolor[HTML]{FFFFFF} &
  \cellcolor[HTML]{FFFFFF} &
  \cellcolor[HTML]{FFFFFF} &
  \multicolumn{3}{c}{\cellcolor[HTML]{FFFFFF}$\Delta \textbf{DNS-MOS}$} \\
\rowcolor[HTML]{FFFFFF} 
\multirow{-2}{*}{\cellcolor[HTML]{FFFFFF}\textbf{Model}} &
  \multirow{-2}{*}{\cellcolor[HTML]{FFFFFF}\begin{tabular}[c]{@{}c@{}}$\Delta \textbf{SDR}$\\ \textbf{(dB)}\end{tabular}} &
  \multirow{-2}{*}{\cellcolor[HTML]{FFFFFF}\begin{tabular}[c]{@{}c@{}}$\Delta \textbf{PESQ}$\\ \textbf{(MOS)}\end{tabular}} &
  \multirow{-2}{*}{\cellcolor[HTML]{FFFFFF}\begin{tabular}[c]{@{}c@{}}$\Delta \textbf{eSTOI}$\\ \textbf{(\%)}\end{tabular}} &
  \textbf{BAK} &
  \textbf{OVRL} &
  \textbf{SIG} \\ \midrule
Teacher      & 8.65          & 1.25          & 10.07         & 1.44          & 0.69          & 0.06           \\
Student      & 6.34          & 0.75          & 5.82          & 1.27          & 0.55          & -0.02          \\ \midrule
\rowcolor[HTML]{FFFFFF} 
$\textbf{Distillation}$ &               &               &               &               &               &                \\ \midrule
Output~\cite{response_distill_SE}       & 6.35          & 0.75          & 5.59          & 1.33          & 0.56          & -0.03          \\
$\textbf{G}$~\cite{sp}             & 6.32          & 0.75          & 5.70          & 1.29          & 0.56          & \textbf{-0.02}          \\
$\textbf{G}_{t}$~\cite{cross_similarity}         & 6.50          & \textbf{0.77}          & 5.95          & 1.33          & 0.55          & -0.04          \\
$\textbf{G}_{f}$         & 6.47          & 0.74          & \textbf{6.03} & 1.29          & 0.56          & \textbf{-0.02} \\
$\textbf{G}_{tf}$ (ours)        & \textbf{6.68} & \textbf{0.77} & 5.99          & \textbf{1.36} & \textbf{0.57} & -0.04          \\ \bottomrule
\end{tabular}
}
\end{table}
\endgroup

\subsection{Self-similarity local KD}
\label{ssec:local_results}

Table~\ref{tab:local_distill_table} shows the efficacy of local similarity-based one-step KD approaches when training student models from scratch.
Using teacher output as $\L_{KD}$ in Eq.~\ref{eq:L} or $\textbf{G}$ similarity~\cite{sp} does not improve, or even deteriorates the student performance. $\textbf{G}_t$ similarity proposed in~\cite{cross_similarity} provides 0.16 dB SDR improvement over the student alone, alongside the best PESQ score. Our proposed time-frequency similarity calculation method $\textbf{G}_{tf}$ outperforms $\textbf{G}_t$ by doubling its SDR improvement (+0.34 dB, w.r.t.~the student alone) and increasing all other scores. This suggests that increasing the granularity of the similarity matrix in the $\L_{KD}^{local}$ calculation facilitates the KD process and overall improves the performance of the distilled student model.

\begingroup
\setlength{\tabcolsep}{4pt} 
\begin{table}[]
\caption{Two-step KD. \textbf{Step~1} - Student pre-training using only $\L_{KD}$ ($\gamma=1$) or no pre-training (None). \textbf{Step~2 } - $\L_{PSA}$: student training with only PSA loss ($\gamma=0$; supervised), $\textbf{G}_{tf}$: Loss from Eq.~\ref{eq:L} using $\textbf{G}_{tf}$-based $\L_{KD}$ and $\gamma=0.5$ (best from Table~\ref{tab:local_distill_table}).}\vspace{-4.5pt}
\resizebox{\columnwidth}{!}{
\centering
\small
\begin{tabular}{@{}cccccccc@{}}
\toprule
\rowcolor[HTML]{FFFFFF} 
\multicolumn{2}{c}{\cellcolor[HTML]{FFFFFF}} &
  \cellcolor[HTML]{FFFFFF} &
  \cellcolor[HTML]{FFFFFF} &
  \cellcolor[HTML]{FFFFFF} &
  \multicolumn{3}{c}{\cellcolor[HTML]{FFFFFF}$\Delta \textbf{DNS-MOS}$} \\
\rowcolor[HTML]{FFFFFF} 
\multicolumn{2}{c}{\multirow{-2}{*}{\cellcolor[HTML]{FFFFFF}\textbf{Model}}} &
  \multirow{-2}{*}{\cellcolor[HTML]{FFFFFF}\begin{tabular}[c]{@{}c@{}}$\Delta \textbf{SDR}$\\ \textbf{(dB)}\end{tabular}} &
  \multirow{-2}{*}{\cellcolor[HTML]{FFFFFF}\begin{tabular}[c]{@{}c@{}}$\Delta \textbf{PESQ}$\\ \textbf{(MOS)}\end{tabular}} &
  \multirow{-2}{*}{\cellcolor[HTML]{FFFFFF}\begin{tabular}[c]{@{}c@{}}$\Delta \textbf{eSTOI}$\\ \textbf{(\%)}\end{tabular}} &
  \textbf{BAK} &
  \textbf{OVRL} &
  \textbf{SIG} \\ \midrule
\multicolumn{2}{c}{Teacher}                 & 8.65          & 1.25          & 10.07         & 1.44          & 0.69          & 0.06           \\
\multicolumn{2}{c}{Student}      & 6.34          & 0.75          & 5.82          & 1.27          & 0.55          & -0.02          \\ 
\midrule
\rowcolor[HTML]{FFFFFF} 
\textbf{Step 1}               & \textbf{Step 2} &               &               &               &               &               &                \\ \midrule
\multirow{-1}{*}{None}       & $\textbf{G}_{tf}$        & 6.68          & 0.77          & 5.99          & \textbf{1.36} & 0.57          & -0.04          \\ \midrule
                             & $\L_{PSA}$            & 6.46          & 0.78          & 6.07          & 1.29          & 0.56          & -0.02          \\
\multirow{-2}{*}{\begin{tabular}[c]{@{}c@{}}$\textbf{G}^{i\rightarrow j}_{t}$\end{tabular}}  & $\textbf{G}_{tf}$        & 6.54          & 0.78          & 5.88          & 1.33          & 0.56          & -0.04          \\ \midrule
                             & $\L_{PSA}$            & 6.54          & 0.79          & 5.87          & 1.33          & 0.57          & -0.02          \\
\multirow{-2}{*}{\begin{tabular}[c]{@{}c@{}}$\textbf{G}^{i\rightarrow j}_{tf}$\end{tabular}} & $\textbf{G}_{tf}$        & 6.76          & 0.80          & 6.06          & 1.33          & 0.57          & -0.03          \\ \midrule
                             & $\L_{PSA}$            & \textbf{6.77} & \textbf{0.81} & \textbf{6.38} & 1.34          & \textbf{0.59} & \textbf{-0.01} \\
\multirow{-2}{*}{$\textbf{G}_{tf}$}      & $\textbf{G}_{tf}$        & 6.75          & 0.80          & 6.34          & 1.32          & 0.57          & -0.02          \\ \bottomrule
\end{tabular}
}
\label{tab:init_table}
\end{table}
\endgroup

\subsection{Two-step KD}
\label{ssec:init_results}

Table~\ref{tab:init_table} presents the results of the proposed two-step distillation process described in Section~\ref{sec:two_step}. We find that using time-preserving flow matrices $\textbf{G}^{i\rightarrow j}_{t}$as the $\L_{KD}$ pre-training objective (Step 1) yields comparable or worse performance than using local similarity $\textbf{G}_{tf}$ with no pre-training. However, changing the pre-training objective to $\textbf{G}^{i\rightarrow j}_{tf}$, which captures interactions between latent representations in greater detail, yields improvement across nearly all the metrics when paired with $\textbf{G}_{tf}$-based KD as Step 2. Most interestingly, pre-training the student with $\textbf{G}_{tf}$ criterion and continuing with only the supervised loss $\L_{PSA}$ provides substantial improvements across all the metrics, especially SDR (+0.44 dB, w.r.t. the student alone) and eSTOI (+0.56\%), suggesting improved intelligibility.

\begin{figure}[t]
    \centering
    \includegraphics[width=\columnwidth]{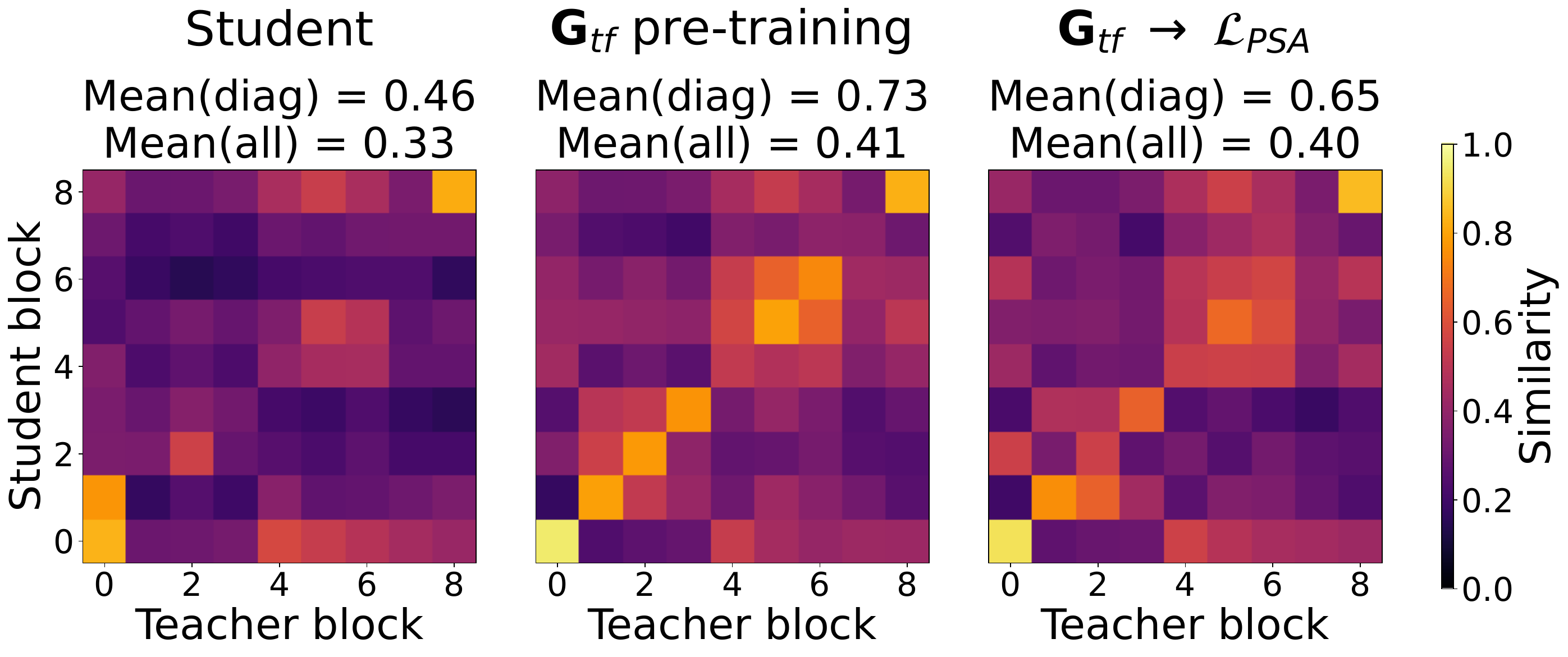}
    \caption{Block-wise CKA similarity between students and teacher networks, averaged over the MS-DNS test set. 
    \textit{Mean(diag)} and \textit{Mean(all)} denote the average similarity for the corresponding blocks (diagonal) or all the block combinations, respectively.}
    \label{fig:cka_fig}
\end{figure}

We further investigate our best two-stage KD approach by performing Central Kernel Alignment (CKA)~\cite{cka} analysis. In principle, CKA allows us to compare the similarity between activation patterns across different models in response to a set of inputs. We use the entire evaluation dataset to probe the models and compute CKA similarities for each pair of layers (averaged over all the audio clips). Fig.~\ref{fig:cka_fig}-left presents CKA similarity between the teacher and student trained independently. Fig.~\ref{fig:cka_fig}-middle compares the teacher to the student pre-trained with $\textbf{G}_{tf}$ criterion (only Step 1). As expected, the first step alone increases the similarity between the corresponding teacher and student layers (diagonal). Finally, Fig.~\ref{fig:cka_fig}-right shows the best student from Table~\ref{tab:init_table}, namely Step 1: $\textbf{G}_{tf}$ $\L_{KD}$-only pre-training, and Step 2: fully-supervised $\L_{PSA}$. The overall similarity to the teacher decreases but remains much higher than for the student trained independently. 
This suggests that a brief pre-training distillation ($\gamma=1$, no $\L_{PSA}$) allows the student to develop its unique solution starting from strong prior knowledge inherited from the teacher.

\begingroup
\setlength{\tabcolsep}{4pt} 
\begin{table}[]
\caption{Evaluation of the two-step KD approach on the MS-DNS dataset remixed at fixed SNRs. \textit{Proposed}: two-step KD using $\textbf{G}_{tf}$ pre-training followed by fully-supervised training (best in Table~\ref{tab:init_table}).}\vspace{-4.5pt}
\label{tab:snr_eval_table}
\resizebox{\columnwidth}{!}{
\centering
\begin{tabular}{@{}cclcccccc@{}}
\toprule
\rowcolor[HTML]{FFFFFF} 
\cellcolor[HTML]{FFFFFF} &
  \multicolumn{2}{c}{\cellcolor[HTML]{FFFFFF}} &
  \cellcolor[HTML]{FFFFFF} &
  \cellcolor[HTML]{FFFFFF} &
  \cellcolor[HTML]{FFFFFF} &
  \multicolumn{3}{c}{\cellcolor[HTML]{FFFFFF}$\Delta \textbf{DNS-MOS}$} \\
\rowcolor[HTML]{FFFFFF} 
\multirow{-2}{*}{\cellcolor[HTML]{FFFFFF}\begin{tabular}[c]{@{}c@{}}\textbf{SNR (dB)}\end{tabular}} &
  \multicolumn{2}{c}{\multirow{-2}{*}{\cellcolor[HTML]{FFFFFF}\textbf{Model}}} &
  \multirow{-2}{*}{\cellcolor[HTML]{FFFFFF}\begin{tabular}[c]{@{}c@{}}$\Delta \textbf{SDR}$\\ \textbf{(dB)}\end{tabular}} &
  \multirow{-2}{*}{\cellcolor[HTML]{FFFFFF}\begin{tabular}[c]{@{}c@{}}$\Delta \textbf{PESQ}$\\ \textbf{(MOS)}\end{tabular}} &
  \multirow{-2}{*}{\cellcolor[HTML]{FFFFFF}\begin{tabular}[c]{@{}c@{}}$\Delta \textbf{eSTOI}$\\ \textbf{(\%)}\end{tabular}} &
  \textbf{BAK} &
  \textbf{OVRL} &
  \textbf{SIG} \\ \midrule
 &
  \multicolumn{2}{c}{Teacher} &
  14.05 &
  0.62 &
  19.12 &
  2.16 &
  1.02 &
  0.64 \\
 &
  \multicolumn{2}{c}{Student} &
  10.82 &
  0.30 &
  10.07 &
  1.86 &
  0.79 &
  \textbf{0.51} \\
\multirow{-3}{*}{- 5} &
  \multicolumn{2}{c}{Proposed} &
  \textbf{11.73} &
  \textbf{0.35} &
  \textbf{11.61} &
  \textbf{1.98} &
  \textbf{0.81} &
  0.47 \\ \midrule
 &
  \multicolumn{2}{c}{Teacher} &
  12.30 &
  0.92 &
  17.83 &
  1.99 &
  0.98 &
  0.40 \\
 &
  \multicolumn{2}{c}{Student} &
  9.65 &
  0.49 &
  10.56 &
  1.75 &
  0.75 &
  \textbf{0.26} \\
\multirow{-3}{*}{0} &
  \multicolumn{2}{c}{Proposed} &
  \textbf{10.23} &
  \textbf{0.56} &
  \textbf{11.51} &
  \textbf{1.84} &
  \textbf{0.79} &
  0.25 \\ \midrule
 &
  \multicolumn{2}{c}{Teacher} &
  10.27 &
  1.21 &
  13.98 &
  1.65 &
  0.78 &
  0.02 \\
 &
  \multicolumn{2}{c}{Student} &
  7.97 &
  0.69 &
  8.58 &
  1.44 &
  0.59 &
  -0.10 \\
\multirow{-3}{*}{5} &
  \multicolumn{2}{c}{Proposed} &
  \textbf{8.43} &
  \textbf{0.76} &
  \textbf{9.32} &
  \textbf{1.51} &
  \textbf{0.62} &
  \textbf{-0.09} \\ \bottomrule
\end{tabular}
}
\end{table}
\endgroup

\subsection{Impact of the student model size and mixture SNR}

The MS-DNS evaluation dataset consists of relatively high SNRs between 0 and 19 dB (mean 9.07 dB). To assess the SNR-dependent benefit of the proposed two-step KD approach, we remix the entire evaluation set to obtain mixtures of the same speech and noise clips but at fixed SNRs of -5, 0 and 5 dB. In Table.~\ref{tab:snr_eval_table}, we observe the inverse relationship between the benefit of our approach and SNR of the noisy mixtures. In particular, for -5 dB SNR mixtures, our KD approach improves student performance by approximately 1 dB SDR, 1.5\% eSTOI, and 0.1 DNS-MOS BAK. This is an important observation as tiny SE models (here, 3.3\% teacher size) tend to exhibit the most significant performance drop in the low-SNR cases, compared to their larger counterparts~\cite{fedorov2020tinylstms}.

In Table~\ref{tab:model_size_table} we showcase the efficacy of the proposed KD framework for various student sizes using the same teacher. We observe that the smaller the downstream model, the larger benefit our KD method provides over the student trained alone. In particular, for the 30k-parameter student ($\sim$1.5\% teacher size), the improvements are the largest with over 1 dB SDR, 0.1 PESQ, and nearly~2\% eSTOI. For model sizes above 200k ($\sim$15\% teacher size), the improvements start to plateau. These findings indicate that our method provides the largest performance boost for the most resource-constrained cases, usually deemed as the most challenging~\cite{chen2019deep,fedorov2020tinylstms,stamenovic2021weight}.

\section{Conclusions}
This work proposes a novel two-step KD protocol for distilling tiny, causal SE models. No previous KD work has investigated this class of embedded-scale SE. Our framework consists of two distinct steps: 1. Distilling the student model using only KD objective using only our proposed fine-grained self-similarity matrix $\textbf{G}_{tf}$ for computing distillation loss; 2. Training the model obtained in Step 1 via supervised loss. Our results show that tiny SE models distilled in this fashion perform better than KD methods utilizing weighted loss between supervised and distillation objectives. Our experimental evaluation shows that the proposed two-step KD provides the largest benefits for low-SNR mixtures and smaller student models. 
Future work should explore integrating the proposed two-step KD with pruning and/or quantization to achieve SE models of even lower complexity and apply the method to other audio-to-audio problems such as source separation, bandwidth extension, or signal improvement.

\begingroup
\setlength{\tabcolsep}{2pt} 
\begin{table}[]
\caption{Impact of the student model size on the two-step KD performance. OPS:~number of operations per frame at inference time. \textit{Proposed}: two-step KD using $\textbf{G}_{tf}$ pre-training followed by fully-supervised training (best in Table~\ref{tab:init_table}).}\vspace{-4.5pt}
\label{tab:model_size_table}
\resizebox{\columnwidth}{!}{
\centering
\small
\begin{tabular}{@{}ccccccccc@{}}
\toprule
\rowcolor[HTML]{FFFFFF} 
\cellcolor[HTML]{FFFFFF} &
  \cellcolor[HTML]{FFFFFF} &
  \cellcolor[HTML]{FFFFFF} &
  \cellcolor[HTML]{FFFFFF} &
  \cellcolor[HTML]{FFFFFF} &
  \cellcolor[HTML]{FFFFFF} &
  \multicolumn{1}{c}{\cellcolor[HTML]{FFFFFF}$\Delta \textbf{DNS-MOS}$} \\
\rowcolor[HTML]{FFFFFF} 
\multirow{-2}{*}{\cellcolor[HTML]{FFFFFF}\textbf{Model}} &
  \multirow{-2}{*}{\cellcolor[HTML]{FFFFFF}\begin{tabular}[c]{@{}c@{}}\textbf{Params / OPS}\\ \textbf{(M)}\end{tabular}} &
  \multirow{-2}{*}{\cellcolor[HTML]{FFFFFF}\begin{tabular}[c]{@{}c@{}}$\Delta \textbf{SDR}$\\ \textbf{(dB)}\end{tabular}} &
  \multirow{-2}{*}{\cellcolor[HTML]{FFFFFF}\begin{tabular}[c]{@{}c@{}}$\Delta \textbf{PESQ}$\\ \textbf{(MOS)}\end{tabular}} &
  \multirow{-2}{*}{\cellcolor[HTML]{FFFFFF}\begin{tabular}[c]{@{}c@{}}$\Delta \textbf{eSTOI}$\\ \textbf{(\%)}\end{tabular}} &
  \textbf{BAK} &
  \textbf{OVRL} &
  \textbf{SIG} \\ \midrule
Teacher &
  1.9 / 13.34 &
  8.65 &
  1.25 &
  10.07 &
  1.44 &
  0.69 &
  0.06 \\ \midrule\midrule
\rowcolor[HTML]{FFFFFF} 
Student &
   &
  4.42 &
  0.50 &
  2.59 &
  \textbf{1.21} &
  0.47 &
  -0.07 \\
Proposed &
  \multirow{-2}{*}{0.03 / 0.42} &
  \textbf{5.52} &
  \textbf{0.61} &
  \textbf{4.55} &
  1.18 &
  0.47 &
  \textbf{-0.05} \\ \midrule
Student &
   &
  6.34 &
  0.75 &
  5.82 &
  1.27 &
  0.55 &
  -0.02 \\
Proposed &
  \multirow{-2}{*}{0.06 / 0.84} &
  \textbf{6.77} &
  \textbf{0.81} &
  \textbf{6.38} &
  \textbf{1.34} &
  \textbf{0.59} &
  \textbf{-0.01} \\ \midrule
Student &
   &
  7.24 &
  0.93 &
  7.53 &
  1.38 &
  0.62 &
  0.00 \\
Proposed &
  \multirow{-2}{*}{0.24 / 2.48} &
  \textbf{7.60} &
  \textbf{0.97} &
  \textbf{7.71} &
  \textbf{1.41} &
  \textbf{0.64} &
  \textbf{0.01} \\ \midrule
Student & 
   &
  7.51 &
  0.99 &
  7.97 &
  \textbf{1.39} &
  0.63 &
  0.01 \\
Proposed &
  \multirow{-2}{*}{0.35 / 3.08} &
  \textbf{7.54} &
  \textbf{1.01} &
  \textbf{8.22} &
  1.38 &
  \textbf{0.64} &
  \textbf{0.02} \\ \bottomrule
\end{tabular}
}
\end{table}
\endgroup

\vfill\pagebreak


\bibliographystyle{IEEEbib}
\bibliography{strings,refs}

\end{document}